\documentclass[12pt,twoside]{article}
\usepackage{fleqn,espcrc1,epsfig}

\usepackage{graphicx}
\usepackage[figuresright]{rotating}

\newlength{\ziffer}
\newcommand{\0}{\settowidth{\ziffer}{0}\hspace*{\ziffer}}

\newcommand{\TeV}{\,\mbox{Te\kern-0.2exV}}
\newcommand{\GeV}{\,\mbox{Ge\kern-0.2exV}}
\newcommand{\MeV}{\,\mbox{Me\kern-0.2exV}}
\newcommand{\keV}{\,\mbox{ke\kern-0.2exV}}
\newcommand{\eV}{\,\mbox{e\kern-0.2exV}}

\newcommand{\bea}{\pagebreak[3]\begin{samepage}\begin{eqnarray}}
\newcommand{\eea}{\end{eqnarray}\end{samepage}\pagebreak[3]}
\newcommand{\beq}{\begin{equation}}
\newcommand{\eeq}{\end{equation}}

\newcommand{\ee}{$e^+e^-$}
\newcommand{\as}{$\alpha_s$}
\newcommand{\Bmax}{B_{\mathrm{max}}}

\newcommand{\ecm}{E_{\mathrm{cm}}}

\newcommand{\asb}{$\alpha_0$}

\newcommand{\abb}{Fig.~\ref}
\newcommand{\fig}{\abb}
\newcommand{\tab}{Tab.~\ref}

\newcommand{\oas}{$\cal O$($\alpha_s^2$)}
\newcommand{\oass}{$\cal O$($\alpha_s^3$)}

\newcommand{\jetset}{{\sc Jetset}}

\newcommand{\herwig}{{\sc Herwig}}

\newcommand{\event}{{\sc Event}}

\newcommand{\lep}{{\sc LEP}}
\newcommand{\delphi}{{\sc Delphi}}
\newcommand{\alephh}{{\sc Aleph}}
\newcommand{\opal}{{\sc Opal}}
\newcommand{\ldrei}{{\sc L3}}
\newcommand{\sld}{{\sc SLD}}

\newcommand{\jade}{{\sc Jade}}

\newcommand{\captionstil}{\it}
\newcommand{\danielscaption}[1]{\caption{\captionstil{#1}}}

\newcommand{\vierplotbeschreibunga}
{
Dargestellt sind die Daten (nach Untergrundsubtraktion) 
mit statistischem Fehler, die Simulation nichtradiativer 
$q\bar{q}$-Ereignisse und der WW-Untergrund.
Für die Verteilungen sind in der oberen Kurve die Akzeptanzkorrekturen
wie folgt dargestellt: 
Die Korrektur durch Selektion hadronischer Ereignisse $C_{\mathrm{QCD}}$
als durchgezogene Linie, \dotfill
}
\newcommand{\vierplotbeschreibungb}
{
\ldots dazu (für die Energien mit WW-Untergrund) die Korrektur aufgrund der Schnitte gegen
WW-Ereignisse $C_{\mathrm{WW}}$ als gestrichelte Linie und das Verhältnis der
Daten vor und nach der WW-Subtraktion als gepunktete Linie.
Für die drei höchsten Energien zeigt die unterste Kurve das Produkt aus
Reinheit und Effizienz der angewendeten WW-Unterdrückung.
\\~~
}

\newcommand{\vierplots}[5]
{
\begin{figure}[ht]
 \begin{center}
  \vspace*{-2.5cm}
  \unitlength1cm
  \begin{minipage}[t]{7.1cm}
   \mbox{\epsfig{file=figs/#2,width=8cm}}
  \end{minipage}
  \begin{minipage}[t]{7.1cm}
   \mbox{\epsfig{file=figs/#3,width=8cm}}
  \end{minipage}
 \end{center}

\vspace*{-3.8cm}
 \begin{center}
  \unitlength1cm
 \begin{minipage}[t]{7.1cm}
   \ifthenelse{\equal{#4}{none.eps}}
              {~}
              {\mbox{\epsfig{file=figs/#4,width=8cm}}}
  \end{minipage}
  \begin{minipage}[t]{7.1cm}
   \mbox{\epsfig{file=figs/#5,width=8cm}}
  \end{minipage}
 \end{center}
\vspace*{-1.5cm}
      \danielscaption{ #1 }
\end{figure}
}

\newcommand{\vierplotsb}[5]
{
\begin{figure}[ht]
 \begin{center}
  \vspace*{-2.5cm}
  \unitlength1cm
  \begin{minipage}[t]{7.1cm}
   \mbox{\epsfig{file=figs/#2,width=8cm}}
  \end{minipage}
  \begin{minipage}[t]{7.1cm}
   \mbox{\epsfig{file=figs/#3,width=8cm}}
  \end{minipage}
 \end{center}

\vspace*{-3.8cm}
 \begin{center}
  \unitlength1cm
 \begin{minipage}[t]{7.1cm}
   \mbox{\epsfig{file=figs/#4,width=8cm}}
  \end{minipage}
  \begin{minipage}[t]{7.1cm}
   \mbox{\epsfig{file=figs/#5,width=8cm}}
  \end{minipage}
 \end{center}
\vspace*{-1.5cm}
\vskip\abovecaptionskip 
{\captionstil{ #1 }}
\end{figure}
}
\newcommand{\achtplots}[3]
{  \clearpage
   \ifthenelse{\isodd{\value{page}}}{~\newpage}{}
   \vierplots{#1 \vierplotbeschreibunga}
             {#2_89gev_00f28pp18h.eps}
             {#2_91gev_00f28pp18h.eps}
             {#3.eps}
             {#2_93gev_00f28pp18h.eps}
             
   \vierplotsb{\vierplotbeschreibungb}
             {#2_133gev_00f28pp18h.eps}
             {#2_161gev_01f28pp18hd.eps}
             {#2_172gev_01f28pp18hd.eps}
             {#2_183gev_01f28pp18hd.eps}
 }

\title{Review on $\alpha_s$ at LEP}

\author{Daniel Wicke\address{Fachbereich Physik, Bergische Universit\"at-GH,
        Gau{\ss}str. 20, D-42097 Wuppertal; wicke@cern.ch\\ 
        }}

\begin{document}

\maketitle

\begin{abstract}
To measure the strong coupling $\alpha_s$ from event shape observables
two ingredients are necessary. A perturbative prediction containing
the dependence of observables on $\alpha_s$ and a description of the
hadronisation process to match the perturbative prediction with the
hadronic data. 

As perturbative prediction  ${\cal O}(\alpha_s^2)$,
NLLA and combined calculations are available. Beside the well known
Monte-Carlo based models also analytical predictions, so called power
corrections, exist to describe the hadronisation. 
Advantages and disadvantages of the different resulting methods for
determining the strong coupling and its energy dependence
will be discussed, the newest \delphi\ results will be presented, and an
overview of the \lep\ results will be included.
\end{abstract}

\section{INTRODUCTION}
The theoretical description of hadron production in \ee-annihilation
consists of three parts. The first  part is based on the
fundamentals of the standard model: Feynman diagrams are used to
(perturbatively)  
calculate the electroweak process of \ee-annihilation 
and the evolution of partons under the strong interaction.
At some stage the partons must be combined
to become hadrons. This hadronisation process cannot be described
by perturbation theory and thus builds a second (non-perturbative) part.
Finally the decay of unstable hadrons, which can be described by kinematics
using experimentally measured decay rates, need to be included before the
prediction can be confronted with data.

Several different predictions exist for the two first 
parts. 
Calculations of the evolution of quarks and gluons are available in 
fixed order \oas, recently some observables became available in \oass, and in
the next to leading log approximation (NLLA), 
which resums large logarithms to
all orders of \as.
When combining \oas\ with NLLA matching ambiguities occur leading to
even more competing predictions.

To describe the second part (the hadronisation) usually generator based models are used.
The most reliable of these models 
are the Lund string fragmentation, implemented in \jetset, and the 
Cluster fragmentation, implemented in \herwig.
Recently analytical predictions, so called power corrections, 
became available to describe the influence of 
hadronisation on  event shape observables.

Given the many different possibilities to determine \as\ 
it was necessary to restrict this 
review 
due to time and space limitation. 
I've chosen to review the
analyses using event shape observables to determine \as,
as this field currently is very active at \lep.

In the following I will first discuss the standard method for measuring \as\
from event shapes, which uses \oas, NLLA or the combined \oas+NLLA prediction in
combination with generator based hadronisation models.
In the second part the energy dependence of \as\ will be discussed introducing
the alternative analytic description of hadronisation by power corrections.
Finally first results obtained using the new \oass\ predictions are presented.
In all three sections \delphi-analyses shall serve as showcases.

\section{CONSISTENT DETERMINATION OF THE STRONG COUPLING}
The original motivation of repeating an \as-measurement with \lep1 data 
in \delphi\ was to
include the event orientation, given e.g.\ by the polar angle of the Thrust
axis $\theta_T$, into  the fit.
Second order coefficients including such an event orientation became available
in \oas\ through the \event2\ program by Catani and 
Seymour~\cite{Catani:1997vz}.
Using the fully reprocessed LEP1 data, event shapes were measured   
very precisely in eight bins of~$\theta_T$. The small systematic 
errors result
not only from the good quality of the final data 
reprocessing, but also from the
fact that all detector corrections in this measurement were naturally
calculated for each $\theta_T$-bin separately.

The second order prediction for such distributions now contains second order 
coefficients  $A$ and
$B$ which depend on the observables value itself and
on $\theta_T$:
\beq \frac{1}{N}\frac{\mathrm{d}^2N}{\mathrm{d}y\,\mathrm{d}\theta_T}  
  =  A(y,\theta_T)   \frac{\alpha_s(\mu)}{2\pi}+
               \left(A(y,\theta_T)\cdot 2 \pi b_0
                 \ln\left(\frac{\mu^2}{\ecm^2}\right) 
                 + B(y,\theta_T)\right)
                \left(\frac{\alpha_s(\mu)}{2\pi}\right)^2
\label{eq_oas_y_theta}
\eeq
$\mu$ being the renormalisation
scale and $b_0=(33-2N_f)/12\pi$.
Beside \as\ this formula contains
renormalisation scale $x_\mu$ parametrised as 
$x_\mu=\mu^2/\ecm^2$ as a free parameter. 

\begin{figure}
\hfill\mbox{\epsfig{file=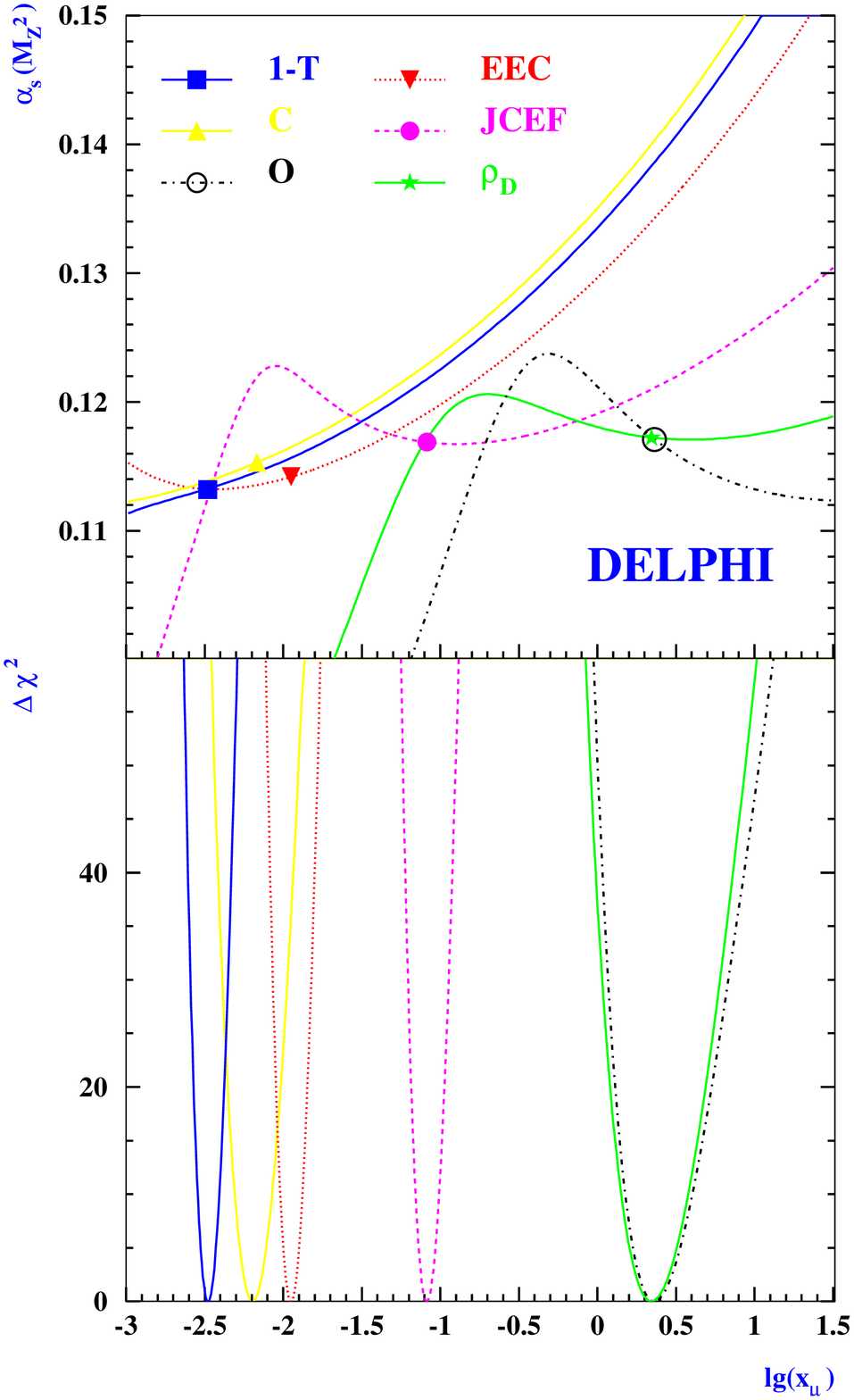,width=5.7cm}\hspace{-1.0cm}
\epsfig{file=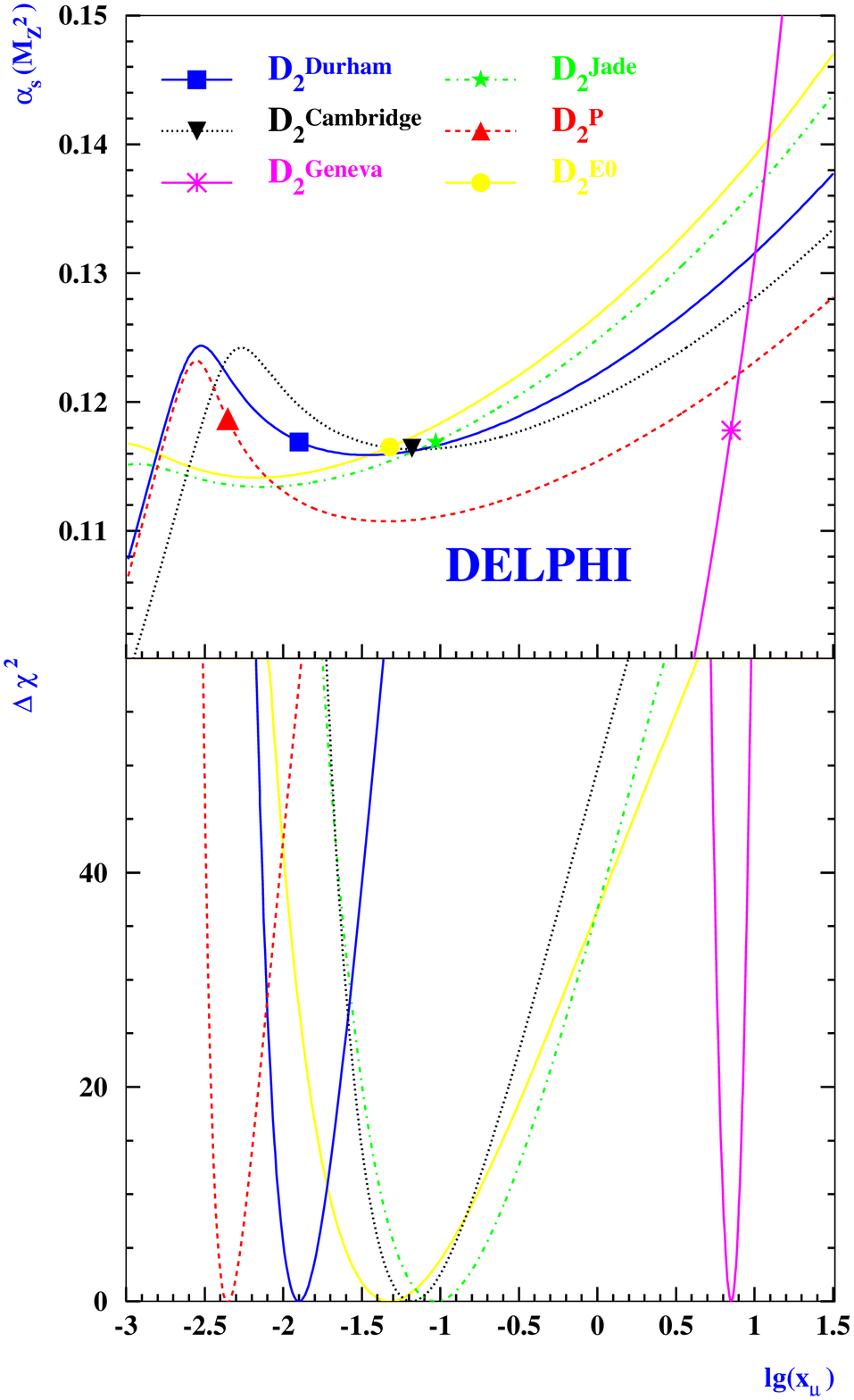,width=5.7cm}\hspace{-1.0cm}
\raisebox{-3mm}[0cm]{\epsfig{file=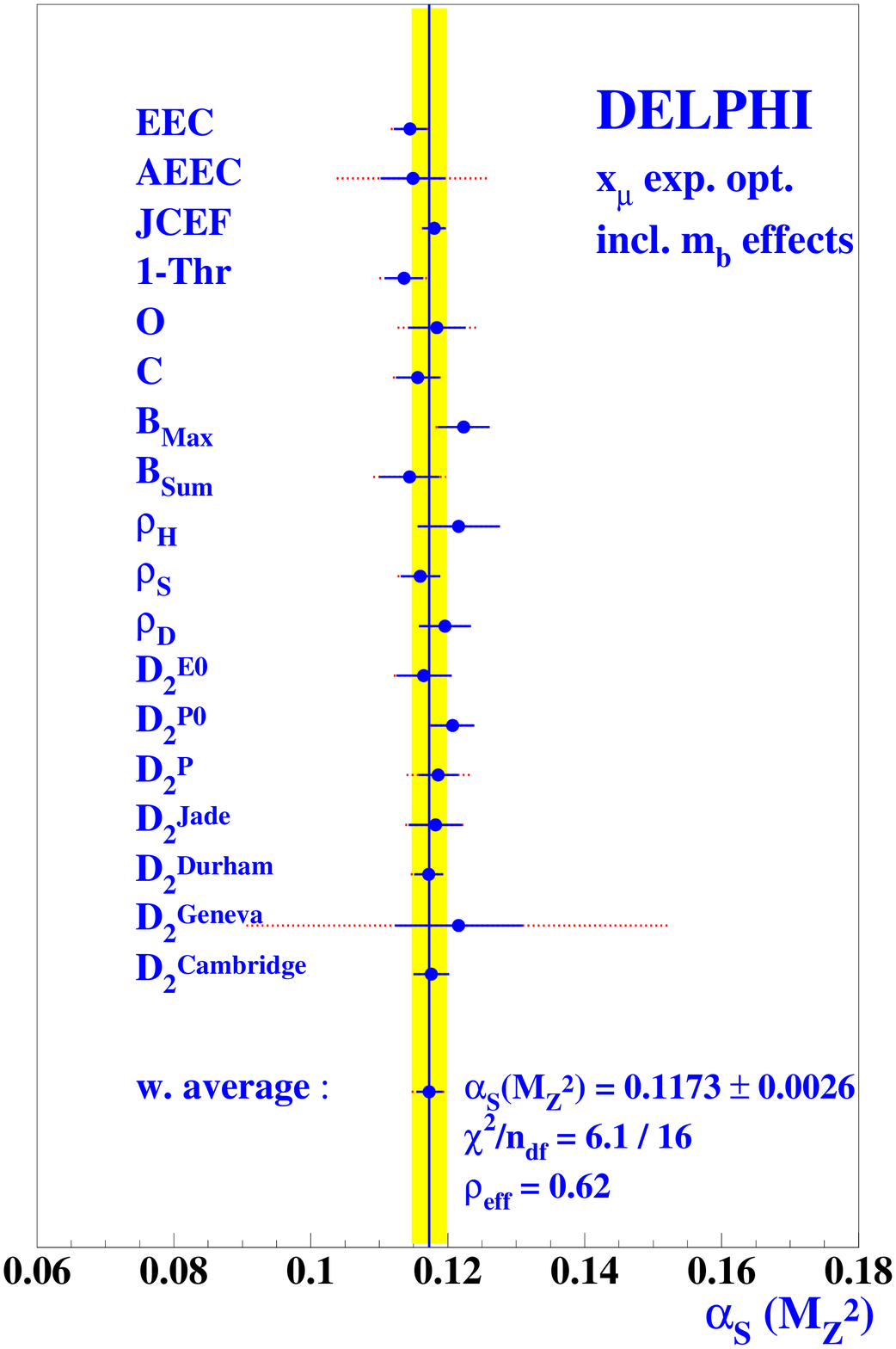,width=6.0cm}}
}
\vspace*{-4ex}
\caption{\label{fig_siggi}%
  Dependence of the resulting \as\ on the renormalisation scale for
  different event shapes (upper parts in left and middle plot) and the
  corresponding $\Delta\chi^2$-values (lower parts). Resulting \as\ values
  for 18
  different event shape observables using experimentally optimised scale
  and including corrections for $b$-quark mass  (right plot). 
}
\end{figure}
As a first step the dependence of the resulting \as-values on this parameter
were investigated. It is found that different  observables aquire largely
different dependencies on $x_\mu$. Also the scales with the optimal
$\chi^2/\mathrm{ndf}$ vary widely (see \fig{fig_siggi} left and middle). 
In spite of this large variation of the
optimal scale, the \as\ results corresponding to the experimentally optimised
scales show a smaller spread  for a large number of different event shapes,
than the results obtained with $x_\mu=1$. In contrast to the results
with scale $1$, optimised scales yield consistent values of \as\ without
assuming any renormalisation scale error (see \fig{fig_siggi} right). 
In addition the scale dependence of \as\ near
the optimised scale is smaller than for $x_\mu=1$, so that the scale variation
between half and twice the chosen value of $x_\mu$ leads to a
smaller scale uncertainty for optimised scales.

Averaging over all 18 investigated observables the final
result~\cite{eps-hep99_1-224} is
\bea
\alpha_s(M_{\mathrm{Z}})&=&0.1228\pm0.0119\quad\mbox{($x_\mu=1$)}  \\
\alpha_s(M_{\mathrm{Z}})&=&0.1173\pm0.0026\quad\mbox{(opt. scales),}
\eea
where for the optimised scales also the $b$-quark mass
corrections are included.
The overall fit quality of these results
is far better for the optimised scales than for $x_\mu=1$, 
as can be seen from the $\Delta\chi^2$ in \fig{fig_siggi}. 
Thus the larger spread of the results with  $x_\mu=1$ can be attributed to a
less stable fit procedure which is caused by the bad agreement between data
and the prediction.

To crosscheck the results the scale dependence was also investigated for
NLLA and combined \oas+NLLA predictions. In contrast to the \oas\ results the
required relative renormalisation scales are close to one and thus there is no
significant 
change compared to the results obtained with $x_\mu=1$. Moreover, because of 
the limited fit range available for NLLA and because of the matching
ambiguities for combined \oas+NLLA predictions the total error for these
methods is larger than for the \oas\ result:
\bea
\alpha_s(M_{\mathrm{Z}})&=&0.116\pm0.006\qquad\mbox{NLLA}\\
\alpha_s(M_{\mathrm{Z}})&=&0.119\pm0.005\qquad\mbox{\oas+NLLA}
\eea
Both results are in good agreement with each other and with the \oas\
results~\cite{eps-hep99_1-224}.

There are two other publication in which optimised scales are used to
determine \as\ at $\ecm=M_{\mathrm{Z}}$: Already in 1992 
\opal~\cite{ZPhysC55_1} states to have found a
clear improvement in the fit quality with optimised scales using 14
observables. They quote 
\bea
\alpha_s(M_{\mathrm{Z}})&=&0.118^{+0.007}_{-0.003}\quad\mbox{(opt. scales),}
\eea
where the error includes a scale variation from the optimised scale upto a
scale of 1.

In 1996 Burrows et. al.~\cite{Burrows:1996vt}
investigated 15 Observables using \sld-data. In contrast to
\opal\ and \delphi\ they found no significant reduction of the spread of
\as-values, though the shift to lower values of \as\ when using optimised
scales  is reproduced:
\bea
\alpha_s(M_{\mathrm{Z}})&=&0.1265\pm0.0076\quad\mbox{($x_\mu=1$)}\\
\alpha_s(M_{\mathrm{Z}})&=&0.1173\pm0.0071\quad\mbox{(opt. scales).}
\eea

In spite of the extra theoretical uncertainties due to matching ambiguities
three of four \lep\ experiments today use the combined \oas+NLLA calculation to
determine their central \as-value
(second error at \ldrei\ is the theoretical component):
\beq
\renewcommand{\arraystretch}{1.15}
\begin{array}{rcll}
\alpha_s(M_{\mathrm{Z}})&=&0.1216\pm0.0039         &\quad\mbox{\alephh~\cite{eps-hep99_1-410}}\\
\alpha_s(M_{\mathrm{Z}})&=&0.1220\pm0.0015\pm0.0060&\quad\mbox{\ldrei~\cite{l3note2414}}\\
\alpha_s(M_{\mathrm{Z}})&=&0.120\0\pm0.006\0&\quad\mbox{\opal~\cite{ZPhysC55_1}.}
\end{array}
\eeq

\section{ENERGY DEPENDENCE}
The increase of beam energy accomplished during the LEP2 programme gives access
to the energy dependence of event shapes and thereby to the energy dependence
of \as.

\subsection{Power Corrections}
Using data at different energies allows also to replace the generator based  
hadronisation models by an analytical ansatz. For mean values
this ansatz (which was developed by Dokshitzer and
Webber~\cite{PhysLettB352_451})  
describes the hadronisation by an additive term:
\beq
\left< f \right> = 
\frac{1}{\sigma_{\mathrm{tot}}}
\int f\frac{\mathrm{d}f}{\mathrm{d}\sigma}\mathrm{d}\sigma =
\left<f_{\mathrm{pert}}\right>+\left<f_{\mathrm{pow}}\right>\quad\mbox{.}
\label{eq_f}
\eeq
The 2nd order perturbative prediction is given by Eq.~(\ref{eq_oas_y_theta})
with coefficients A and B  integrated over $y$ and $\theta_T$. 
The power correction term is falling off like the
inverse centre-of-mass energy and is given by
\beq
\left < f_{pow}\right >  =  c_f \frac{4C_F}{\pi^2} {\cal M}
\frac{\mu_I}{E_{\mathrm{cm}}} 
  \left[{\alpha}_0(\mu_I) - \alpha_s(\mu)
        - \left(b_0 \cdot \log{\frac{\mu^2}{\mu_I^2}} + 
\frac{K}{2\pi} + 2b_0 \right) \alpha_s^2(\mu) 
\right]\nonumber
\label{eq_fpow_dw}
\eeq
\begin{table}[b]
\begin{center}
\renewcommand{\arraystretch}{1.2}
\begin{tabular}{|c|c|c|c|} \hline
Observable    & ${\alpha}_0(2\GeV)$         
                &$\alpha_s(M_{\mathrm{Z}})$ 
                     & $\chi^2/\mathrm{ndf}$ \\ \hline
$\left<\Bmax\right>$
 & $0.407 \pm 0.022 \pm 0.055$
                & $0.117 \pm 0.0012\pm 0.0015$
                        & 8.28/21 \\ 
$\left<1-T\right>$ 
& $0.493 \pm 0.009 \pm 0.006$
                & $0.119 \pm 0.0014\pm 0.0067$
                        & 64.0/32 \\
$\left<M_{\mathrm{h}}^2/E_{\mathrm{vis}}^2\right>$ 
 & $0.545 \pm 0.023 \pm 0.017$
                & $0.120 \pm 0.0020\pm 0.0050$
                        & 8.28/21 \\ 
\hline 
\end{tabular}
\end{center}
\caption{\label{tab_mess_fit}Determination of \asb\ in a combined fit of \asb\
                and \as\ using a large number of
                measurements~\cite{collection_eventshapes} at different
                energies~\cite{eps-hep99_1-144}. 
                The first error is calculated from the fit, the
                second contains the renormalisation scale dependence.} 
\end{table}
where \asb\ is a non-perturbative parameter accounting for the
contributions to the event shape below an infrared matching scale $\mu_I$,
$K=(67/18-\pi^2/6)C_A-5N_f/9$. The Milan factor ${\cal M}$ is set to
1.8, which corresponds to three active flavours in the
non-perturbative region~\cite{NuclPhysB511_396}. 
The observable-dependent coefficient  $c_f$ is 2
and 1 for $f=\left<1-T\right>$   
and $f=\left<M_{\mathrm{h}}^2/E_{\mathrm{vis}}^2\right>$, respectively.
For $\left<\Bmax\right>$ the coefficient is itself energy dependent:
$c_{\left<\Bmax\right>}\sim 1/\sqrt{\alpha_s(\ecm)}$~\cite{Dokshitzer:1998qp}.
The infrared matching scale is set to 2\GeV\ as suggested
by the authors~\cite{PhysLettB352_451}, the renormalisation scale
$\mu$ is set to be equal to ${E_{\mathrm{cm}}}$.
Beside \as\ these formulae contain \asb\ as the only free parameter. In
order to measure \as\ from individual high energy data this
parameter has to be known.

To infer \asb, a combined fit of \as\ and \asb\ to a large set of
measurements at different energies~\cite{collection_eventshapes}
is performed~\cite{PhysLettB456_322,eps-hep99_1-144}. 
For $\ecm\ge M_{\mathrm{Z}}$ only \delphi\ measurements are included in
the fit. 
The resulting values of \asb\ for
$\left<\Bmax\right>$, $\left<1-T\right>$ and
$\left<M_{\mathrm{h}}^2/E_{\mathrm{vis}}^2\right>$
are summarised in \tab{tab_mess_fit}.
Even though the found \asb\ values are experimentally inconsistent, the
universality of \asb\ is not violated because the expected  theoretical
precision allows deviations of upto 20\%.
The \asb\ values are consistent with the corresponding analyses of
\ldrei~\cite{l3note2414}  
and \opal/\jade~\cite{eps-hep99_1-5}.

\begin{figure}[t]
\vspace*{-1cm}
\begin{tabbing}
\hspace{1.6cm}\=\hspace{-1.9cm}%
\epsfig{file=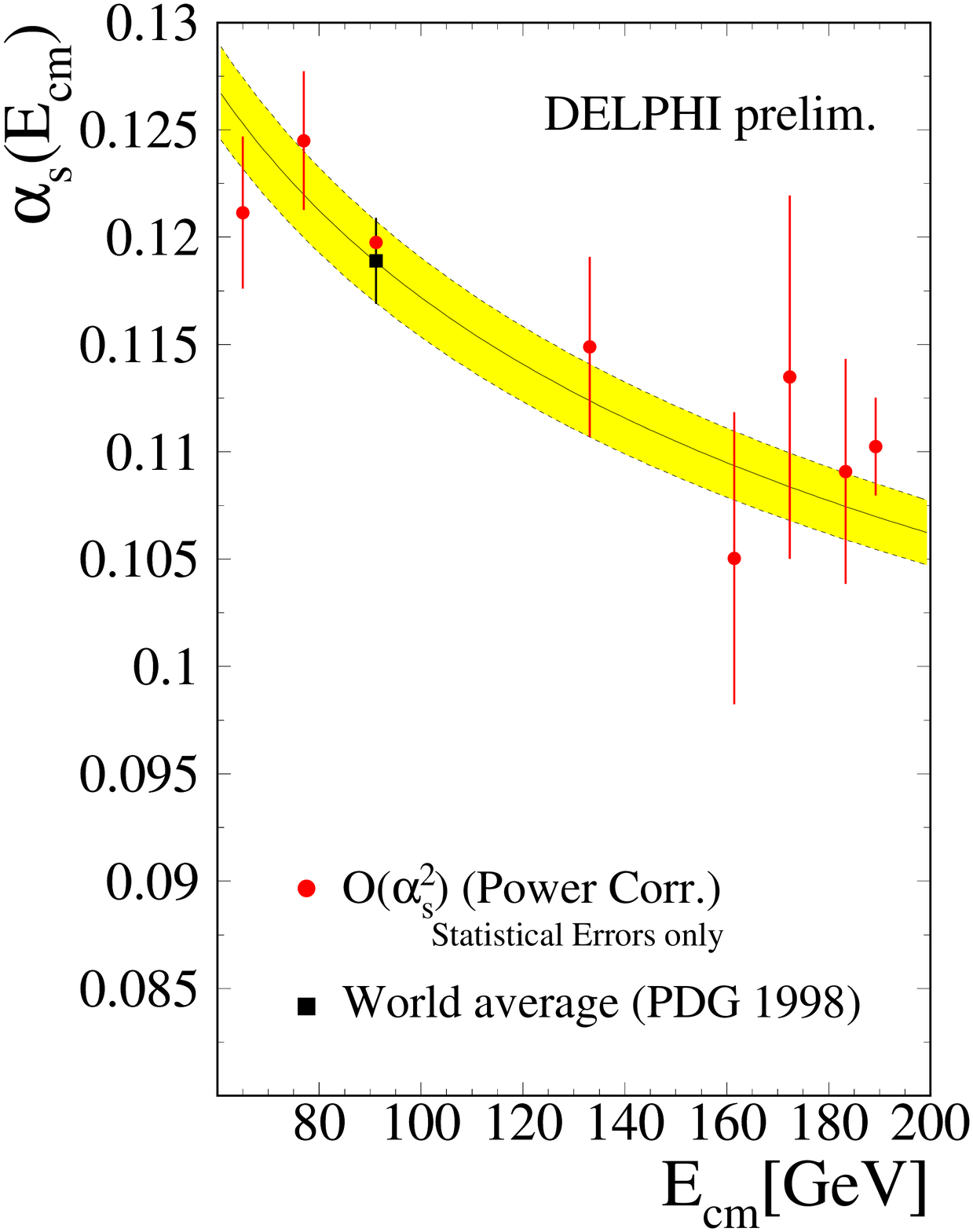,width=5cm}
\=\hspace{-1.4cm}\epsfig{file=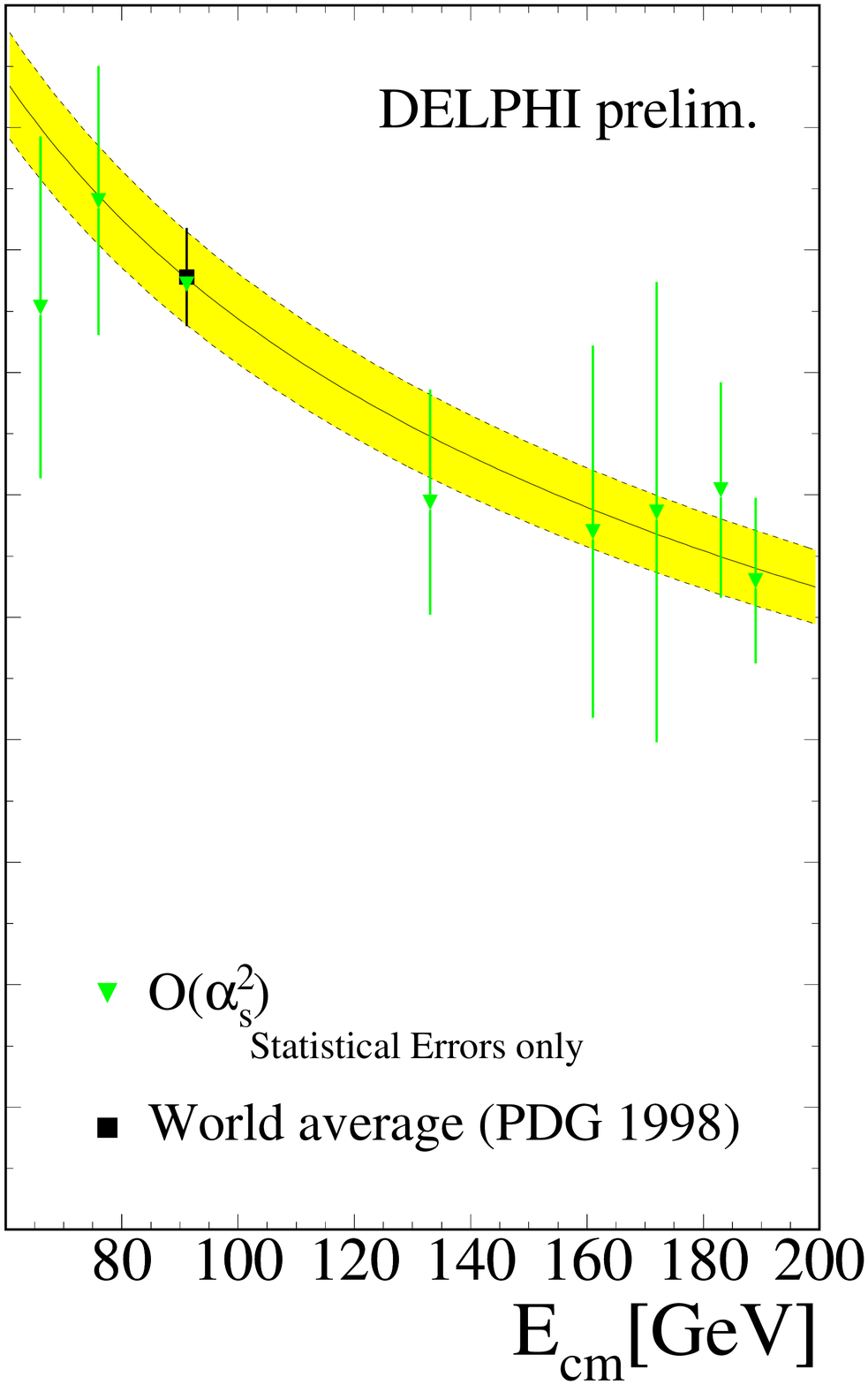,width=5cm}
\=\hspace{-1.4cm}\epsfig{file=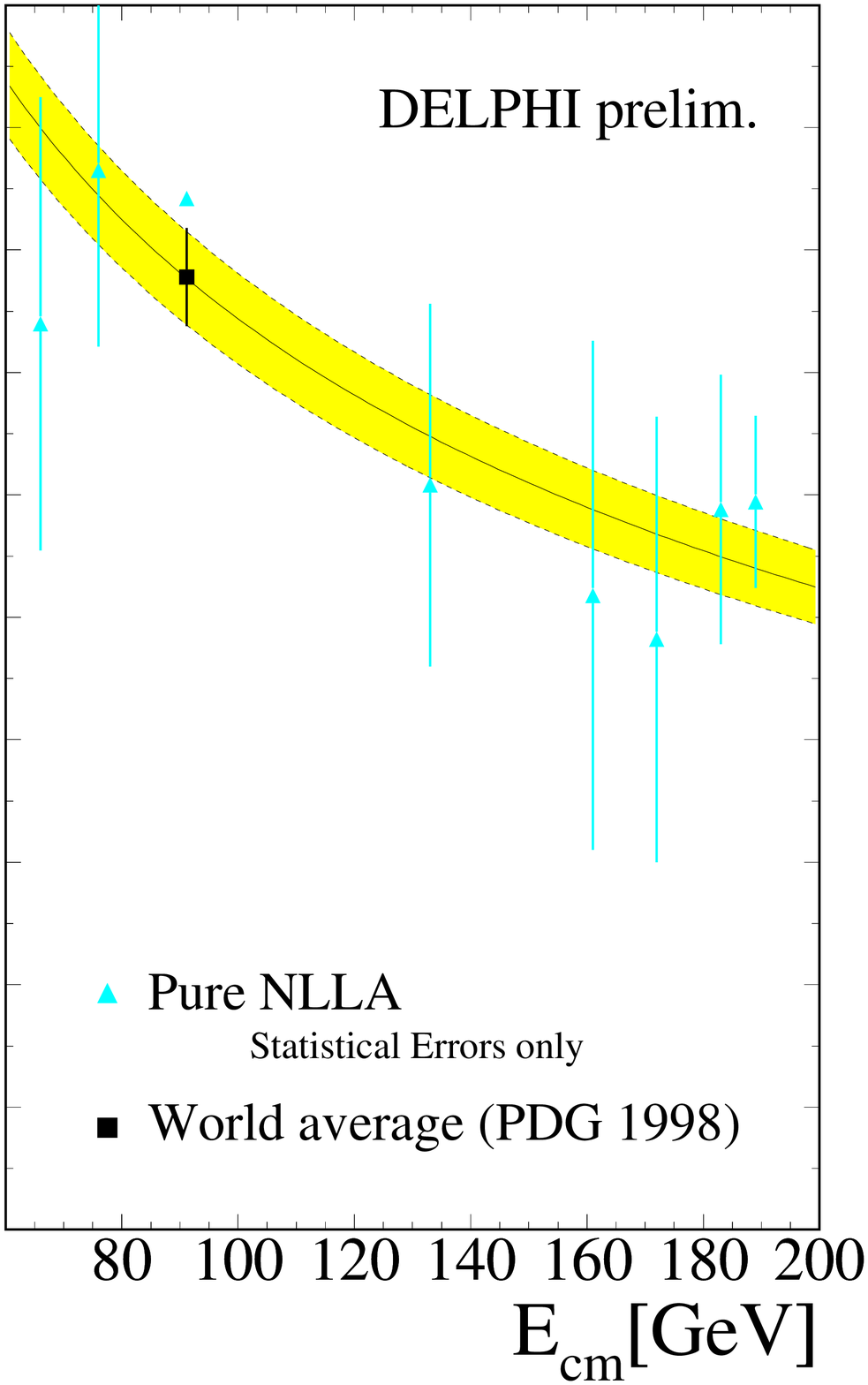,width=5cm}
\=\hspace{-1.4cm}\epsfig{file=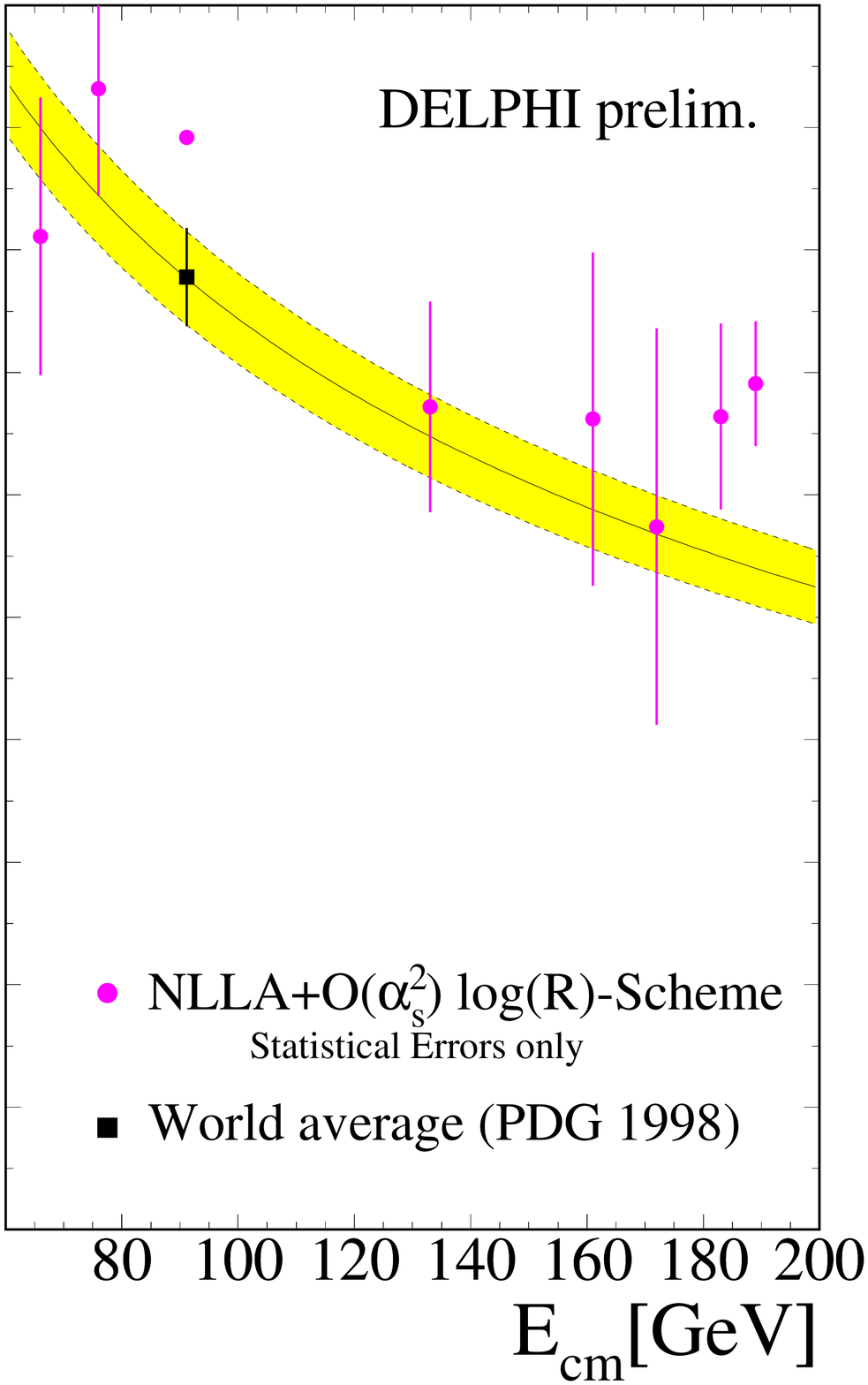,width=5cm}\\
\>Power Corr.\>Monte Carlo based hadronisation models  \\
\> \oas          \> \oas     \> NLLA               \> \oas+NLLA  \\
{\footnotesize$\displaystyle \frac{\mathrm{d}\alpha_s^{-1}}{\mathrm{d}\log\ecm}$}
\>$1.07${\small$\pm0.19\pm0.20$}  \>$1.23${\small$\pm0.31\pm0.19$}  \>$1.41${\small$\pm0.31\pm0.27$}      \>{$1.12${\small$\pm0.22\pm0.18$}}
\normalsize
\end{tabbing}
\vspace*{-1em}
\caption{\label{fig_as_results}Comparison of \as-results obtained by \delphi\ at different
  centre-of-mass energies using Monte Carlo generator based hadronisation
  models as well as power correction. Below each plot the result for the
  logarithmic energy slope of the inverse coupling is given.
  }
\end{figure}

After fixing \asb\ to the values found, \as\ can be calculated
individually for each energy using Eqs. (\ref{eq_f}--\ref{eq_fpow_dw}).
The results for energies between 65\GeV\ and 189\GeV\ 
of this method and of the traditional methods
described in the previous section are compared in \fig{fig_as_results}.
All methods give consistent results.

\subsection{Energy dependence of \boldmath\as}
To measure the energy dependence of \as\ \delphi\ uses the logarithmic energy
slope of the inverse coupling. This quantity is directly proportional to the
Callan-Symanzik $\beta$-function and is independent of \as\ and of $\ecm$ to
first order:
\bea
\frac{\mathrm{d}\alpha_s^{-1}}{\mathrm{d}\log\ecm}
&=&-\frac{1}{\alpha_s^2}\beta_{\alpha_s}
=\frac{\beta_0}{2\pi}+\frac{\beta_1}{4\pi^2}\alpha_s+\ldots 
\simeq 1.27
\eea
The numerical value represents the QCD prediction calculated in second order
for energies between 91\GeV\ and 200\GeV\ using the PDGs world average of \as.
The energy dependence of this derivative in the given range and the
uncertainty of \as\ influence this value by about one unit in the last
digit. 

The result with the smallest systematic error is obtained from \oas+NLLA fits:
\bea
\frac{\mathrm{d}\alpha_s^{-1}}{\mathrm{d}\log\ecm}&=&1.12\pm0.22\pm0.18
\eea
in good agreement with the QCD expectation.

Instead of determining a value for the energy dependence explicitly \ldrei\
checks the running of \as\ by applying a combined fit to all energies assuming
the standard model running to \oass. This yields a
$\chi^2/\mathrm{ndf}=15.4/12$ also 
indicating a good agreement~\cite{l3note2414}.

\section{\oass}
Third order calculations for four-parton final states have become available
recently~\cite{Dixon:1997th,Nagy:1997yn}. 
These calulations can be used to measure \as\ in next to leading
order from event shapes that acquire non-trivial values only for four and more
partons, like
the four-jet-rate $R_4$:
\beq
 R_4=B\left(\frac{\alpha_s(\mu)}{2\pi}\right)^2
    +\left(B\cdot 2 \pi b_0 \log\frac{\mu^2}{E_{\mathrm{cm}}^2}+C\right)
    \left(\frac{\alpha_s(\mu)}{2\pi}\right)^3
\label{eq_oass}
\eeq
Observables of this kind are uncorrelated to the observables discussed so far
which are based on three-jet-like configurations. The reduced number of 
relavant events is partly compensated by the (due to quadratic \as\
dependence) increased sensitivity in Eq.~(\ref{eq_oass}).

\delphi\ investigated four-jets-rates $R_4(y_{\mathrm{cut}})$ at a given
$y_{\mathrm{cut}}$ for the Durham cluster
algorithm~\cite{flagmeyer:montpellier99}. 
It shows good agreement with the prediction for $y_{\mathrm{cut}}>0.002$.
Using $y_{\mathrm{cut}}=0.0025$ the resulting \as\ values are consistent with
the results shown in the previous section, but they have larger statistical
errors. Also the running obtained from this analysis is in good agreement:
\bea
\frac{\mathrm{d}\alpha_s^{-1}}{\mathrm{d}\log\ecm}&=&1.16\pm0.46\quad\mbox{.}
\eea
This new calculation has also been investigated by \alephh, 
but was not yet used to determine the strong coupling.

\section{SUMMARY}
In this talk it was tried to give an overview over the current state of the
art in measuring \as\ from event shape observables at \lep. Beside different
perturbative calculations several hadronisation models (generator based and
analytical ones) exist. 
The previous standard choice for the perturbative calculation \oas+NLLA is
questioned by \delphi\ due to poor data description and
unsatisfactory consistency.

In that sense measurements of \as\ from \lep\ data are 
still in full progress. 
In addition to the method used, special attention is payed to tests of the
running, also new theoretical developements like power corrections and \oass\
gain due recognition. 

\newcommand{\esdcollection}
{  
 ALEPH  Coll., D. Decamp et al. {\em Phys. Lett.} {\bf B284} (1992) 163.\\
 ALEPH Coll., D. Buskulic et al. {\em Z. Phys.} {\bf C55} (1992) 209.\\ 
 AMY  Coll.,  I.H. Park et al. {\em Phys. Rev.} Lett. {\bf 62} (1989) 1713.\\ 
 AMY Coll., Y.K. Li et al. {\em Phys. Rev.} {\bf D41} (1990) 2675. \\ 
 CELLO Coll., H.J. Behrend et al. {\em Z. Phys.} {\bf C44} (1989) 63. \\ 
 HRS Coll., D. Bender et al. {\em Phys. Rev.} {\bf D31} (1985) 1.\\ 
 P.A. Movilla Fernandez, et. al. and the JADE Coll. 
 {\em Eur. Phys. J.} {\bf C1} (1998) 461.\\
 L3 Coll., B. Adeva et al. {\em Z. Phys.} {\bf C55} (1992) 39.\\ 
 Mark II Coll., A. Peterson et al. {\em Phys. Rev.} {\bf D37} (1988) 1. \\ 
 Mark II Coll., S. Bethke et al.   {\em Z. Phys.} {\bf C43} (1989) 325.\\ 
 MARK J Coll.,  D. P. Barber et al. {\em Phys. Rev. Lett.} {\bf 43} (1979) 831.\\ 
 OPAL Coll., P. Acton et al. {\em  Z. Phys.} {\bf C59} (1993) 1.\\ 
 PLUTO Coll., C. Berger et al. {\em Z. Phys.} {\bf C12} (1982) 297. \\ 
 SLD Coll., K. Abe et al. {\em Phys. Rev.} {\bf D51} (1995) 962.\\ 
 TASSO Coll., W. Braunschweig et al. {\em Phys.} Lett. {\bf B214} (1988) 293.\\ 
 TASSO Coll., W. Braunschweig et al. {\em Z. Phys.} {\bf C45} (1989) 11.\\ 
 TASSO Coll., W. Braunschweig et al. {\em Z. Phys.} {\bf C47} (1990) 187.\\ 
 TOPAZ Coll.,  I. Adachi et al. {\em Phys. Lett.} {\bf B227} (1989) 495.\\ 
 TOPAZ Coll., K. Nagai et al. {\em Phys. Lett.} {\bf B278} (1992) 506. \\ 
 TOPAZ Coll., Y. Ohnishi et al. {\em Phys. Lett.} {\bf B313} (1993) 475.
}
\bibliographystyle{utphysnt}
\bibliography{QCD}

\providecommand{\href}[2]{#2}\begingroup\begin{thebibliography}{10}

\bibitem{Catani:1997vz}
S.~Catani and M.~H. Seymour {\em Nucl. Phys.} {\bf B485} (1997) 291--419,
  \href{http://xxx.lanl.gov/abs/hep-ph/9605323}{{\tt hep-ph/9605323}}.

\bibitem{eps-hep99_1-224}
{\sc DELPHI},  EPS-HEP99 \# 1\_224, 1999.

\bibitem{ZPhysC55_1}
{\sc OPAL}, P.~D. Acton {\em et.~al.} {\em Z. Phys.} {\bf C55} (1992) 1.

\bibitem{Burrows:1996vt}
P.~N. Burrows, H.~Masuda, D.~Muller, and Y.~Ohnishi {\em Phys. Lett.} {\bf
  B382} (1996) 157, \href{http://xxx.lanl.gov/abs/hep-ph/9602210}{{\tt
  hep-ph/9602210}}.

\bibitem{eps-hep99_1-410}
{\sc ALEPH}  {ALEPH 99-023, EPS-HEP99 \# 1\_410}, 1999.

\bibitem{l3note2414}
{\sc L3},  {L3 Note 2414, EPS-HEP99 \# 1\_279}, 1999.

\bibitem{PhysLettB352_451}
Y.~L. Dokshitzer and B.~R. Webber {\em Phys. Lett.} {\bf B352} (1995) 451.

\bibitem{NuclPhysB511_396}
Y.~L. Dokshitzer, A.~Lucenti, G.~Marchesini, and G.~P. Salam {\em Nucl. Phys}
  {\bf B511} (1997) 396, \href{http://xxx.lanl.gov/abs/hep-ph/9707532}{{\tt
  hep-ph/9707532}}.

\bibitem{Dokshitzer:1998qp}
Y.~L. Dokshitzer, G.~Marchesini, and G.~P. Salam
  \href{http://xxx.lanl.gov/abs/hep-ph/9812487}{{\tt hep-ph/9812487}}.

\bibitem{collection_eventshapes}
\esdcollection.

\bibitem{PhysLettB456_322}
{DELPHI Coll., P. Abreu {\em et.~al.}} {\em Phys. Lett.} {\bf B456} (1999) 322.

\bibitem{eps-hep99_1-144}
{\sc DELPHI},  EPS-HEP99 \# 1\_144, 1999.

\bibitem{eps-hep99_1-5}
{\sc OPAL} and {\sc JADE}  EPS-HEP99 \# 1\_5, 1999.

\bibitem{Dixon:1997th}
L.~Dixon and A.~Signer {\em Phys. Rev.} {\bf D56} (1997) 4031--4038,
  \href{http://xxx.lanl.gov/abs/hep-ph/9706285}{{\tt hep-ph/9706285}}.

\bibitem{Nagy:1997yn}
Z.~Nagy and Z.~Trocsanyi {\em Phys. Rev. Lett.} {\bf 79} (1997) 3604--3607,
  \href{http://xxx.lanl.gov/abs/hep-ph/9707309}{{\tt hep-ph/9707309}}.

\bibitem{flagmeyer:montpellier99}
U.~Flagmeyer Talk given at the {QCD} 99 conference held in {M}ontpellier, July
  1999, to be published in Nucl. Phys. (Proc. Suppl.), 1999.

\end{thebibliography}\endgroup

\end{document}